\documentclass{mem}
\usepackage{natbib}\usepackage{txfonts}\usepackage{balance}
\usepackage{graphicx}
\usepackage[a4paper,breaklinks]{hyperref}
\idline{75}{282}
\begin{document}
\def\teff{$T\rm_{eff }$}
\def\kms{$\mathrm {km s}^{-1}$}

\title{
Gamma-ray lines in modern astrophysics
}

   \subtitle{}

\author{
Fiona H. \,Panther\inst{1}}

\institute{
$^1$School of Science, University of New South Wales Canberra, Australian Defence Force Academy, Canberra 2612, Australia
\email{f.panther@adfa.edu.au}
}

\authorrunning{Panther}

\titlerunning{Gamma-ray lines in modern astrophysics}

\abstract{Gamma-ray astronomy provides a direct window into the most violent, dynamic processes in the Universe. MeV gamma-ray astronomy in particular allows us to directly observe the process of chemical enrichment of the interstellar medium (ISM) through the decay of radioactive isotopes synthesized by stars and compact objects during their lives, in their death throes and after their deaths. Moreover, gamma-ray lines such as the positron annihilation line can give us a unique view into the propagation of cosmic rays in the Galaxy, as well as hints about unusual transient phenomena that may be responsible for producing the positrons. With modern astrophysics moving forward into a new era of multimessenger astrophysics, the important contributions of MeV gamma-ray astronomy must not be overlooked. I briefly outline some of the important contributions gamma-ray spectroscopy with \textit{INTEGRAL} has made to our understanding of the universe, and the future perspectives for this dynamic field of astrophysics.
}
\maketitle{}

\section{Introduction}
One of the key questions that drives modern astrophysics is 'how are the elements formed by stars and recycled through galaxies?'. This is a question we can begin to answer into with the observation of gamma-ray lines produced in the decay of radioactive isotopes synthesized by the death-throes of stars. Depending on the lifetimes of these radioisotopes, we can gain immediate insight into the conditions which forge them (e.g. short-lived isotopes with half-lives of tens of days that originate in the decay chains $^{56}$Ni). On longer timescales, gamma-rays from the radioactive decay of isotopes with half-lives of tens of year reveal the evolution of supernova remnants and the enrichment of the ISM with intermediate mass elements (e.g. $^{44}$Ti decay). Finally, the dynamics of galaxies, which are driven by feedback from massive stars, can be traced via the decay of $^{26}$Al and $^{60}$Fe.\\
Furthermore, all three of these processes can be probed using information gleaned from the observation of positron annihilation, as several of these isotopes are known $\beta^+$ emitters. Identifying the origin of Galactic positrons is a long-standing problem in Galactic and high-energy astrophysics (see \cite{Prantzos11, Panther18a} for detailed reviews), however understanding the origin of the Galactic positrons can help astrophysicists better understand not only chemical enrichment and nucleosynthesis, but also cosmic-ray propagation and Galactic structure.\\
The \textit{INTEGRAL} mission has now reached maturity in the context of gamma-ray line astronomy: a deep understanding of the data reduction process has been achieved \citep{Diehl2017} and observations have moved from being a demonstration of novel technology to providing a robust platform on which both theorists and observers can work to reveal the processes that drive the evolution of our Universe: chemical enrichment, nucleosynthesis, and astroparticle physics. We must now take stock of what we have learned from the first 17 years of the mission, and both theorists and observers must look ahead to future discoveries with \textit{INTEGRAL} and its proposed successors such as \textit{AMEGO}.\\
\section{Supernovae at early times}
On January 22 2014, the closest supernova of any type since 2004 and the closest thermonuclear supernova observed in decades was discovered serendipidously by \cite{Fossey14}. The explosion date is thought to be 14 January, UT 14.75, with 0.21 d uncertainty \citep{Zheng2014}. It was recognized as a type Ia explosion from early spectra \citep{Cao2014} and occurred at a distance of only $3.3\,\mathrm{Mpc}$. As it occurred in a region of the host galaxy (M82) which was highly obscured by dust, the observations were subject to significant reddening \citep{Goobar} with corresponding difficulties to uncover the intrinsic supernova brightness and optical spectra with great precision.\\
Gamma ray observations are not impeded by dust reddening. As the SN occurred so close to Earth, it was possible for INTEGRAL/SPI to make a number of observations of the supernova. Detection of the gamma-ray lines associated with the decay of $^{56}$Co were made by \cite{Churazov14}, \citep{Diehl14}, and \citep{Churazov15} in the months following the supernova explosion.\\
The flux of $^{56}$Co gamma rays was used to infer the $^{56}$Ni mass synthesized in the explosion, determined to be $M_\mathrm{Ni} = 0.6\pm 0.1\,\mathrm{M_\odot}$ by \cite{Churazov14}, and $M_\mathrm{Ni} = 0.49\pm 0.09\,\mathrm{M_\odot}$ \citep{Diehl14}. The differences in the derived Ni masses can be attributed to the differing methods of analysis to determine the Co line fluxes. These observations were the first incontrovertible and direct evidence that thermonuclear supernovae are powered by the decay of the $^{56}$Ni decay chain \citep{Churazov14}, although interpretations of the explosion geometry differ.\\
Future observatories such as AMEGO will have greater sensitivity to the gamma-ray lines produced by the $^{56}$Ni decay chain to greater distances, and thus in the coming decades we can expect a large sample of SNe Ia gamma-ray spectra and lightcurves to be amassed \citep{AMEGONuc}. This will enable statistical analysis to gain a clearer understanding into the progenitor configurations and explosion mechanisms that power these SNe.\\
\section{Supernova Remnants \& $^{44}$Ti}
Massive stars are usually thought of as the main source of the radioisotope $^{44}$Ti. This isotope is formed through explosive nucleosynthesis during the core collapse supernova that occurs at the end of an $>8\,\mathrm{M_\odot}$ star's life. The synthesis of $^{44}$Ti occurs in the alpha-rich freezeout phase of nuclear statistical equilibrium \citep{Thielemann96}.\\
One method to directly investigate the $^{44}$Ti yields of core collapse supernovae is to observe the x-ray and gamma-ray decay lines of the isotope in Milky Way supernova remnants. Observations of the $^{44}$Ti decay lines have been made in the CCSN remnant Cassiopoeia A \citep[Cas A,][]{Iyudin94}.  Spatially-resolved spectroscopic x-ray analyses of the $^{44}$Ti ejecta have been carried out to determine the total initial mass of $^{44}$Ti produced in the explosion, and its velocity structure. In \cite{Grefenstette16}, an initial $^{44}$Ti mass of $1.54\pm0.21\times10^{-4}\,\mathrm{M_\odot}$ was found using the x-ray line flux. In comparison, masses of $1.5\pm0.4\times10^{-4}\,\mathrm{M_\odot}$ and $2.4\pm0.9\times10^{-4}\,\mathrm{M_\odot}$ of $^{44}$Ti, respectively were found using the $78\,\mathrm{keV}$ x-ray line and the $1157\,\mathrm{keV}$ gamma ray line in \cite{Siegert15}.\\
It is possible to also derive the yield of $^{44}$Ti in SN1987A, the closest SN for almost 500 years. Late time observations of SN1987A's lightcurve, which declined at a slower rate than predicted by the decay of $^{56}$Co alone, generate an estimated yield of $^{44}$Ti in the explosion to be $\sim 0.5 \times 10^{-4}\,\mathrm{M_\odot}$, lower than the yield estimates from the x-ray and gamma ray flux observed in Cas A \citep{Seitenzahl2014}. Direct measurements of $^{44}$Ti emission lines from the remnant of SN1987A suggest a larger $^{44}$Ti yield of $3.1\pm0.8\times 10^{-4}\,\mathrm{M_\odot}$ \citep{Grebenev12, Boggs15}.\\
However, there is a dearth of SN remnants which emit the characteristic lines that indicate the presence of $^{44}$Ti. As the Milky Way CCSN rate is estimated to be about $1/\mathrm{century}$, one would expect to see a number of CCSN remnants bright in these decay lines \citep{The06}. Their absence indicates that the typical yield of $^{44}$Ti in CCSNe may be lower than suggested by observations of Cas A and 1987A.\\
However, there is compelling evidence that there may be an additional Galactic source of $^{44}$Ti. The observed abundance of $^{44}$Ca (the daughter nucleus of $^{44}$Ti) in pre-solar grains, relative to $^{56}$Fe suggests that the CCSN rate, which is predicted to be roughly constant across the lifetime of the Galaxy, indicates that even CCSNe producing $^{44}$Ca at $~10^{-4}\,\mathrm{M_\odot}$ per event cannot supply enough $^{44}$Ca \citep{Timmes96}. This hints that there may be another source of $^{44}$Ti in the Galaxy. Such a source would be much rarer, and produce $^{44}$Ti in much greater quantities. This is a natural explanation for the absence of observed SNRs with $^{44}$Ti decay lines \citep{Crocker17}.\\
Two subtypes of thermonuclear supernovae have been directly implicated in the production of $^{44}$Ti. The first is SNe of the subtype SN2005E-like (henceforth 05E-like). A handful of these subluminous and spectroscopically peculiar supernovae have been observed in external galaxies. Their rate is extremely uncertain, with estimates suggesting as low as 2\% of the cosmological SNe Ia rate to as high as 90\% of the SNe Ia rate \citep{Frohmaier2018}. Their rates are primarily uncertain due to low number statistics. Their faint and fast-declining nature means they may be missed in cosmological SN surveys \citep{Lunnan17}.\\
Another subtype of thermonuclear supernovae is a much more compelling candidate as a source of $^{44}$Ti: SN1991bg-like SNe. 91bg-like SNe events share several key features with their normal SNe Ia cousins, and are clearly thermonuclear supernovae. Specifically, they lack any indication of hydrogen and helium in their spectra while also exhibiting strong Si\thinspace {\sc ii}\,in absorption \citep{Fillipenko1997}. Moreover, absorption features in their spectra near maximum light indicate the presence of a number of intermediate mass elements (IME) including silicon, magnesium, calcium, sulphur and oxygen which is consistent with these events belonging to the class of thermonuclear transients \citep{Fillipenko1997}. These supernovae are known to occur predominantly in old stellar populations \citep{Perets10}, and observational constraints on their delay time distribution (Panther et al. 2019 in prep) are consistent with them not only supplying sufficient $^{44}$Ti to explain the missing $^{44}$Ca in presolar grains, but also to explain the origin of up to 90 per cent of Galactic positrons \citep{Crocker17}.\\
\section{Galactic Evolution in Gamma-Rays}
$^{26}$Al is a $\beta^+$ unstable radioisotope of Aluminium with a half-life of $\sim 7.3\times 10^5\,\mathrm{yr}$. It is synthesised by both massive stars (stars with zero-age main sequence masses $>2\,\mathrm{M_\odot}$) and asymptotic giant branch stars (AGB stars) during hydrostatic burning in the hydrogen burning shell of the star \citep{Prantzos96}, and also explosively in the C-Ne-O layers of the stellar interior during supernova explosions \citep{Limongi06}.\\
Stellar winds and the terminal supernova explosions of these stars eject $^{26}$Al into the interstellar medium. $^{26}$Al can be traced via the $1.8\,\mathrm{MeV}$ gamma ray photons emitted when the nucleus decays via $\beta^+$ emission. Thus one can trace the dynamics of material expelled by young, massive stars using the $1.8\,\mathrm{MeV}$ gamma ray line in the Galaxy \citep{Diehl95, Diehl06, Kretschmer13}: Because massive stars live fast and die young, with lifetimes of only a few Myr, $^{26}$Al is known to trace regions of active star formation. In particular, observations by COMPTEL/\textit{CGRO} by \cite{Knoedelseder99} show $^{26}$Al gamma ray emission coincident with nearby OB associations (groups of young, hot, massive stars with spectral types O and B)  Scorpius-Centaurus, Orion and Cygnus. More recent observations of $^{26}$Al gamma rays made by SPI/\textit{INTEGRAL} reveal the kinematic structure of the emission \citep{Kretschmer13}. These observations suggest that $^{26}$Al collects at the edges of superbubbles - large H{\sc{II}}-filled cavities with radii $\sim\mathrm{kpc}$ carved out of the ISM by stellar winds and supernova explosions - in the Galactic disk. These observations have begun to capture the attention of the galaxy evolution community, and works such as \cite{Fujimoto18} have performed simulations in which particles representing the $^{26}$Al are traced to produce synthetic observations of Milky Way-like galaxies in gamma-rays from the nuclear decay of this isotope. These simulations ultimately pave the way for future avenues in which galactic dynamics can be studied at gamma-ray wavelengths.\\
\section{Positron Annihilation in the Milky Way}
The Milky Way hosts the annihilation of $\sim$5$\times 10^{43}$ positrons each second \citep{Siegert16}. The annihilation of positrons is detected indirectly through measurements of gamma rays and is characterised by a strong emission line centered at $511\,\mathrm{keV}$, the rest mass energy of the positron (or electron). Positron annihilation in the Milky Way was first detected by balloon-bourne spectrometers in the early 1970s \citep{Johnson72}: a notable excess of emission at $\sim 0.5\,\mathrm{MeV}$ was observed to be concentrated toward the center of the Galaxy. However, the spatial resolution of such instruments was poor. The most recent observations with SPI/\textit{INTEGRAL} \citep{Knoedelseder05, Weidenspointner08, Siegert16} allow detailed morphological models of positron annihilation gamma rays to be constructed.\\
The most recent morphological models of positron annihilation in the Galactic bulge are described in \cite{Siegert16}, where the emission is modelled as the superposition of two two-dimensional Gaussians. \cite{skinner2015, Siegert16} also describe emission from an extended thick disk, an observation highly dependent on the assumed spatial template. The more robust observation of positron annihilation in the Galactic bulge is the focus of this work, and historically the high surface brightness of positron annihilation gamma rays in this region and the high absolute positron annihilation rate - $\sim2\times10^{43}\,\mathrm{e^+\,s^{-1}}$ - have been difficult to explain. This is because most putative positrons sources are concentrated in regions of star formation in the Galactic disk \citep{Prantzos11, RKL79}, and those associated with the older stellar population of the Galactic bulge have positron yields that are currently not well constrained, such as microquasars \citep{Siegertmicroquasars} and peculiar supernovae \citep{Crocker17}. Others employ exotic physics such as the deexcitation or annihilation of dark matter \citep[e.g.][]{Finkbeiner2007, Boehm09}. Furthermore, large-scale transport into the Galactic bulge by the Milky Way's nuclear outflow has been effectively ruled out \citep{Panther17} and diffusion of positrons over large distances is strongly constrained \citep{Jean09, Martin2012,Alexis14}. Now, one of the great challenges in explaining the origin of the Galactic positrons lies in the fact that positron production is ubiquitous in astrophysical environments. Consequently, the problem for theorists tackling the problem is not which source could produce them, but rather to determine which sources may make a dominant contribution to the positron production rate while also replicating the distribution of the annihilation radiation and the spectral characteristics.\\
The value in taking on these challenges cannot be understated, as it will require us to bring together our understanding of stellar nucleosynthesis, cosmic ray propagation, galactic dynamics, stellar population synthesis, the behavior of matter and gravity in the extreme environments close to black holes and the fundamental particle physics of Dark Matter, if the positrons do indeed have an exotic origin. Furthermore, understanding the distribution of the positron annihilation signal has significantly challenged observers to refine the techniques used to interpret the data obtained by coded-mask instruments.\\
To unveil the origin of the Galactic positrons will require significant interdisciplinary study by both observers and theoreticians. As observers look forward to future gamma-ray telescope missions, we must consider how we plan our observations to better constrain the distribution of the positron annihilation radiation, in particular the scale height and existence of the disk component. For theoreticians, understanding the origin of Galactic positrons in the context of both Galactic chemical evolution and cosmic ray propagation is a fascinating multimessenger astrophysics problem with wide-reaching implications for understanding the evolution of not only the Milky Way but galaxies in general.
\section{Conclusions}
MeV astronomy gives us a unique window into the processes that drive the evolution of the cosmos - the production and recycling of the elements through the stellar life cycle. The discoveries of \textit{INTEGRAL} and its predecessors have already shown the importance of MeV telescopes to gain insight into supernovae and stellar end-products, their remnants, and Galactic dynamics. Proposed future missions such as \textit{AMEGO} promise to provide more observations of gamma-ray lines in our Galaxy and beyond. Combining the greater sensitivity of these future missions with our developing understanding of the complex backgrounds associated with MeV telescopes and the process of data deconvolution, MeV astronomy will continue to provide answers to some of the key questions that drive modern astrophysics.
\begin{acknowledgements}
FHP thanks the SOC of the 12th INTEGRAL conference, particularly Carlo Ferringo, for their generous support. FHP also thanks Roland Diehl and Thomas Siegert for ongoing stimulating discussions on the latest observational gamma-ray line measurements, and Roland Crocker and Ivo Seitenzahl for their ongoing support with regards to theoretical work.
\end{acknowledgements}

\bibliographystyle{aa}
\bibliography{procbib.bib}

\end{document}